\documentstyle[aps,preprint,graphicx]{revtex}

\tightenlines

\begin{document}
\title{Reconciliation of the  
Measurement of Parity-Nonconservation in Cs 
with the Standard Model
}
\author{A. Derevianko} 
\address{Institute for Theoretical Atomic and Molecular Physics\\
Harvard-Smithsonian Center for Astrophysics,
Cambridge, Massachusetts 02138}

\date{\today}
\maketitle

\begin{abstract}
Contributions from the Breit interaction in atomic-structure calculations
account for 1.3$\sigma$ of 
the  previously reported $2.5\sigma$ deviation  from the Standard Model in 
the $^{133}$Cs weak charge [S.C. Bennett and C.E. Wieman,
Phys.\ Rev.\ Lett.\ {\bf 82}, 2484 (1999)].
The updated corrections for the neutron distribution reduce 
the discrepancy further to 1.0$\sigma$.  
The updated value of the weak charge is 
$Q_{\rm W}(^{133}{\rm Cs} ) = -72.65(28)_{\rm expt}(34)_{\rm theor}$. \\
The present analysis is a higher-order extension of previous
calculation [A. Derevianko, E-print physics/0001046].
\end{abstract}

Atomic parity-nonconserving (PNC) experiments combined
with accurate atomic structure calculations provide powerful  constraints
on ``new physics'' beyond the Standard Model of 
elementary particles~\cite{PNC_general}. 
Compared to high-energy experiments or low-energy  scattering experiments, 
atomic single-isotope PNC measurements are uniquely sensitive
to new isovector heavy physics~\cite{Ramsey-Musolf_PRC99}.
Presently, the PNC effect in atoms has been most precisely measured 
by Wieman and co-workers using  $^{133}\mathrm{Cs}$~\cite{Wood_PNC97}.
In 1999, Bennett and Wieman~\cite{BennettWieman_PRL99} 
updated the value of the Cs weak charge by measuring a supporting
quantity, the vector transition polarizability  $\beta$, and
by re-evaluating  the precision of
atomic structure calculations~\cite{Blundell_PNC,Dzuba_PL89} from the 
early 1990s.
The determined weak charge~\cite{BennettWieman_PRL99} differed from the 
prediction~\cite{QW_StdModel}  
of the Standard Model by 2.5 standard deviations $\sigma$.  
The value of the $^{133}\mathrm{Cs}$ weak charge from Ref.~\cite{BennettWieman_PRL99} 
(together with other precision electroweak observables) has been 
employed in numerous articles. 
In particular, recent theoretical investigations~\cite{Casalbuoni_PLB99,Rosner_PRD99} interpret 
this 2.5$\sigma$  deviation as possible evidence for extra neutral vector 
$Z$-bosons. 

The main focus of the two previous {\em ab initio} relativistic 
calculations for the atomic structure of $^{133}$Cs~\cite{Blundell_PNC,Dzuba_PL89} 
was the correlation contribution 
from the residual Coulomb interaction (i.e., beyond Dirac-Hartree-Fock level). 
The purpose of this work is to  evaluate  rigorously contributions from the Breit 
interaction to PNC in  $^{133}$Cs.
The previous calculations either omitted such contributions~\cite{Dzuba_PL89},  
or evaluated them only  partially~\cite{Blundell_PNC}.
The present analysis is a higher-order extension of my recent calculation~\cite{myEprint}. 
It is found that the Breit contribution corrects the 
weak charge by 0.9\%,
reducing the  2.5$\sigma$ deviation from the Standard Model to 1.2$\sigma$.
Including a correction for the neutron density distribution 
in the $^{133}$Cs nucleus further reduces the deviation to $1.0\sigma$.
Thus the result reported here brings the most accurate atomic PNC measurement
to date~\cite{Wood_PNC97} into  substantial agreement with the 
Standard Model.

The Breit interaction~\cite{Johnson_BreitFormulas}
arises due to an exchange of transverse photons
between electrons. Its low-frequency limit, employed here, is given by
\[
   B_{ij} = -\frac{1}{2r_{ij}} 
\left( \alpha_i \cdot \alpha_j + 
(\alpha_i \cdot \hat{r}_{ij} ) (\alpha_j \cdot \hat{r}_{ij} )        \right)
\]
It is convenient to separate 
the second-quantized Breit interaction 
into zero-, one-, and two-body parts normally ordered with respect 
to the core: 
$B = B^{(0)} + B^{(1)} + B^{(2)}$.

The parity-nonconserving  amplitude for the
$6S_{1/2} \rightarrow 7S_{1/2}$ transition in $^{133}\mathrm{Cs}$ can be represented as a sum
over intermediate states $mP_{1/2}$
\begin{eqnarray}
\lefteqn{E_{\mathrm PNC} = \sum_{m} 
\frac{\langle 7S|D|mP_{1/2}\rangle  \langle mP_{1/2} |H_{W}|6S\rangle
}{E_{6S}-E_{mP_{1/2}}}  } \nonumber \\  &+ &
\sum_{m} 
\frac{\langle 7S|H_{W}|mP_{1/2}\rangle  \langle mP_{1/2} |D|6S\rangle
}{E_{7S}-E_{mP_{1/2}}}  
\, .
\label{Eqn_E_PNC}
\end{eqnarray}
Here $D$~\cite{stretch_red} and $H_{\rm W}$ are electric-dipole and weak interaction matrix elements,
and $E_{i}$ are atomic energy levels. 
It is convenient to break the total Breit correction $\delta E_{\mathrm PNC}$ into 
three distinct parts due to corrections in the weak interaction and dipole
matrix elements, and energy denominators, respectively 
\begin{equation}
  \delta E_{\mathrm PNC}  = E_{\mathrm PNC} (\delta H_{\mathrm W}) + E_{\mathrm PNC} (\delta D) +
E_{\mathrm PNC} (\delta E) \, .
\label{Eq_PNC_corrections}
\end{equation}

The overwhelming contribution from parity-violating interactions arises from 
the Hamiltonian
\begin{equation}
 H_{\rm W} = \frac{G_F}{\sqrt{8}} Q_{\rm W} \rho_{\mathrm nuc}(r) \gamma_5 \, ,
\label{Eqn_Hw}
\end{equation}
where $G_F$ is the Fermi constant, $\gamma_5$ is the Dirac matrix,
and $\rho_{\mathrm nuc}(r)$ is the {\em neutron} density distribution.
To be consistent with the previous calculations
the $\rho_{\mathrm nuc}(r)$  is taken to be a {\em proton} Fermi distribution employed in Ref.~\cite{Blundell_PNC}.
The slight difference between the neutron and proton distributions is addressed in the conclusion.
The PNC amplitude is expressed  in units of $10^{-11} i |e| a_0 (-Q_{\rm W}/N)$,
where  $N=78$ is the number of neutrons in the nucleus of $^{133}$Cs. 
In these units the results of past calculations
for  $^{133}\mathrm{Cs}$ are  
$E_{\mathrm PNC} = -0.905$, Ref.~\cite{Blundell_PNC}, and
$E_{\mathrm PNC} = -0.908$, Ref.~\cite{Dzuba_PL89}.
The former value includes a partial Breit contribution $+0.002$, 
and the latter includes none. 
The reference Coulomb-correlated amplitude
\begin{equation}
 E^{C}_{\mathrm PNC} = -0.9075   
\label{Eqn_refVal}
\end{equation}
is determined as an average, 
with the partial Breit contribution removed from the value of Ref.~\cite{Blundell_PNC}.

{\em Hartree-Fock analysis---} 
Before proceeding to the correlated calculations discussed in the
second part of this work, it is worth examining the Breit contribution
to the PNC amplitude at the lowest-order level. 
The conventional Dirac-Hartree-Fock (DHF) equation reads
\begin{equation}
   \left( h_{\rm D} + V_{\rm HF} \right) \phi_i = \varepsilon_i   \phi_i \, ,
\end{equation}
where $h_{D}$ is the Dirac Hamiltonian including the interaction of an electron 
in state $i$ with
a finite-size nucleus. $V_{\rm HF}$ is a mean-field Hartree-Fock potential;
this potential contains direct and exchange Coulomb interactions of electron $i$
with core electrons. A set of DHF equations is solved self-consistently for core orbitals;
valence wavefunctions and energies are determined subsequently by ``freezing''
the core orbitals. 
The Breit-Dirac-Hartree-Fock (BDHF) approximation constitutes the introduction 
of the one-body part of the Breit interaction $B^{(1)}$ into the above DHF equation
\begin{equation}
   \left( h_{\rm D} + \tilde{V}_{\rm HF}  + B^{(1)} \right) \tilde{\phi}_i = 
\tilde{\varepsilon_i} \tilde{\phi_i} \, . \label{Eqn_BDHF}
\end{equation}
Compared to the DHF equations, energies, wave-functions, and the 
Hartree-Fock potential are modified, as designated by tildes. 
This self-consistent
BDHF approximation was used by 
Lindroth {\em et al.}~\cite{Lindroth_Breit} and 
a related iterative
analysis was considered by Johnson {\em et al.}~\cite{Johnson_Breit_PRA88}. 
Both papers point out the importance of the ``relaxation'' effect,
which leads to modification of the Hartree-Fock potential through adjustment of core orbitals. 
In the present work, the relaxation effect is taken into account 
automatically by direct integration of Eq.~(\ref{Eqn_BDHF}). 
 
Most of the Breit contribution to the PNC amplitude can be determined by limiting
the summation over intermediate states in Eq.~(\ref{Eqn_E_PNC}) 
to the two lowest valence $P_{1/2}$ states: $6P_{1/2}$ and $7P_{1/2}$. 
In the DHF approximation
one then finds $E_{\mathrm PNC} =-0.6888$ (90\% of the total value). 
The lowest-order corrections to matrix elements and energy denominators 
calculated as differences between BDHF and DHF values 
are listed in Table~\ref{Tab_HF}. 
The resultant BDHF corrections to $E_{\mathrm PNC}$ are:
\begin{eqnarray}
E_{\mathrm PNC} (\delta H_{\mathrm W}) &=&  0.0022 \; ( 0.32\%) \, , \nonumber \\
E_{\mathrm PNC} (\delta D) &=&    0.0020           \; ( 0.29\% ) \, , \label{Eqn_corr_HF}\\
E_{\mathrm PNC} (\delta E) &=&   -0.0019           \; (-0.28\% )  \, . \nonumber  
\end{eqnarray}
The sum of these three terms leads to $\delta E_{\mathrm PNC} =0.0023$ 
in agreement with the 0.002 correction found by  Blundell {\em et al.}~\cite{Blundell_PNC,Blundell_comment}.
Inclusion of intermediate states beyond $6P_{1/2}$ and $7P_{1/2}$ 
leads to a small additional  modification  to $\delta E_{\mathrm PNC}$ 
of -0.00004.  
Note that if experimental energies
(which effectively include the Breit interaction) are used in the energy 
denominators of Eq.~(\ref{Eqn_E_PNC}),
then the $E_{\mathrm PNC} (\delta E)$ term must be excluded and the total
correction becomes twice as large: $\delta E_{\mathrm PNC} =0.0042$. 

With further examination of the modifications of {\em individual} uncorrelated matrix 
elements presented in Table~\ref{Tab_HF}, one notices the following. \\
(i) Weak interaction matrix elements are each reduced 
in absolute value by 0.3\%, which is directly reflected in a 0.3\% correction to 
the PNC amplitude.\\
(ii) Modification of dipole amplitudes is strongly
nonuniform. There are substantial corrections only to the $6S_{1/2}-7P_{1/2}$ 
(0.5\%) and
$7S_{1/2}-6P_{1/2}$ (0.1\%) matrix elements. 
The large 0.5\% Breit correction to $\langle 6S_{1/2} | D | 7P_{1/2}  \rangle$ 
provides partial resolution to a long-standing 
discrepancy of spectroscopic experiment~\cite{Expt_old} and {\em ab initio} 
calculations~\cite{Dzuba_matels,Blundell_matels,Safronova_AlkSDpT99}.    
The relatively large Breit correction is caused both by an 
accidentally small matrix element and by 
admixture into $\langle 6S_{1/2} | D | 7P_{1/2}  \rangle$ from a 30 times larger  
$7S_{1/2}-7P_{1/2}$ matrix element.  \\
(iii) The largest modification in the energy denominators is
0.1\% for $E_{7S}-E_{6P}$; however, this leads to a 0.3\% correction $E_{\mathrm PNC} (\delta E)$.
As recently emphasized  by Dzuba {\em et al.}~\cite{Dzuba97}, 
such large sensitivity of the resulting PNC amplitude to small variations in individual
atomic properties entering Eq.~(\ref{Eqn_E_PNC}) arises due to a cancellation of relatively large
terms in the sum over states. 

{\em Correlated calculations---}
It is well known that correlations caused by residual Coulomb interactions not included in the
Hartree-Fock equations can lead to substantial modifications of the lowest-order
values. For example, the weak matrix element $\langle 6S_{1/2} | H_{\rm W} | 6P_{1/2}  \rangle$
is increased by a factor of 1.8 by correlations due to residual Coulomb interactions.
It will be shown that the correlations are also  important for a  
proper description of the Breit corrections.

The major correlation effects in atoms appear because of shielding of externally
applied (e.g., electric) fields by core electrons and  an 
additional attraction of the valence electron by an induced dipole moment of the 
core~\cite{thirdorder}. The former effect is described by contributions
beginning  at second order and the latter in  third order of many-body perturbation theory (MBPT).
Since these two effects lead to the dominant contributions in Coulomb-correlated 
calculations, the third-order analysis reported here seems sufficient~\cite{MBPTcomment}. 

\begin{figure}
\centerline{\includegraphics[scale=1.00]{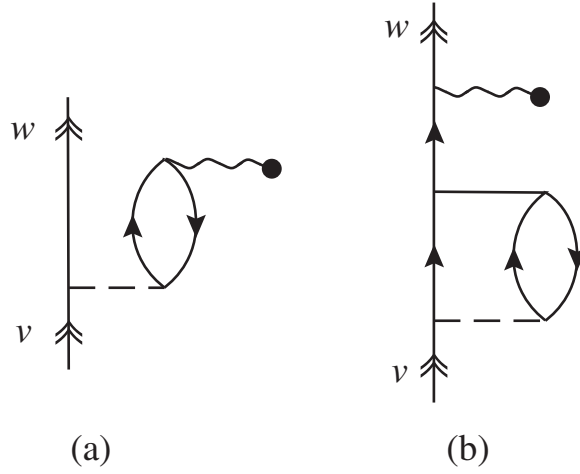}}
\caption{
Sample many-body diagrams included in the calculations.
Dashed (solid) horizontal lines represent the Breit (Coulomb) interaction.
All orbitals are obtained in the Breit-Dirac-Hartree-Fock approximation.
Diagram (a) is one of the contributions in the random-phase approximation,
and diagram (b) is one of  the Brueckner-orbital contributions~\protect\cite{thirdorder}.  
\label{Fig_Breit_diag}}
\end{figure}

MBPT calculations were performed with the two-body Breit interaction $B^{(2)}$ treated on equal footing 
with the residual Coulomb interaction. Sample many-body
diagrams are presented in Fig.~\ref{Fig_Breit_diag}.
To treat the one-body contribution $B^{(1)}$,
an extension of the B-spline basis set technique~\cite{Bsplines} was developed, 
based on the Breit-Dirac-Hartree-Fock (BDHF)
equation~(\ref{Eqn_BDHF}).  Such a formulation made it possible
to handle $B^{(1)}$ and the associated relaxation effect exactly.
Contributions of negative-energy states, discussed for example in 
Ref.~\cite{Savukov_PRL99}, 
were also included and found to be relatively small~\cite{myEprint}. 
Two series of third-order calculations were performed, first with the Breit and Coulomb interactions 
fully included using the BDHF basis set, and second in the DHF basis set without the Breit interaction 
and negative-energy states. The obtained differences are the Breit corrections reported in Table~\ref{Tab_HF}. 

Breit corrections to $^{133}$Cs hy\-per\-fine-struc\-ture mag\-ne\-tic-di\-po\-le 
constants $A$ are discussed first,
since these were considered in the literature previously. 
The correction to hyperfine constants is very sensitive to correlations: 
e.g.,  Ref.~\cite{Blundell_matels} found a numerically insignificant modification for $A_{6S}$, while
Ref.~\cite{Safronova_AlkSDpT99,myEprint} determined the modification to be 
large (-4.64 MHz), 
and the approach reported here yields +4.89 MHz.
In the calculation of Ref.~\cite{Blundell_matels} the correction was determined as a difference of the BDHF
and DHF values, however such approach misses two-body Breit corrections of comparable size. 
In Ref.~\cite{Safronova_AlkSDpT99,myEprint} a second order perturbation analysis
was used for the Breit interaction, but the important relaxation effect discussed earlier was omitted. 
The present calculation incorporates all mentioned diagrams and is also extended to third order. 
Using this same calculational scheme, the corrections to hyperfine constants  for other states 
of $^{133}$Cs are  +1.16 MHz  for $7S_{1/2}$,  -0.51 MHz  for $6P_{1/2}$, and -0.146 MHz for $7P_{1/2}$. 
These corrections improve agreement with experiments for the {\em ab initio} all-order 
Coulomb-correlated calculations~\cite{Blundell_matels} to 0.1\% for all  states except 
$6P_{1/2}$ where the discrepancy becomes 0.5\%.

Examination of the third-order corrections listed in 
Table~\ref{Tab_HF} reveals the significant effect of correlations on the Breit contribution. For example, 
corrections to weak interaction matrix elements become three times larger than those 
in the lowest order.
Compared to hy\-per\-fine-struc\-ture constants 
there is no cancellation of various contributions to the weak 
interaction matrix elements. 
Using third-order matrix elements and second-order energies the 
following {\em ab initio} corrections are determined:
$E_{\mathrm PNC} (\delta H_{\mathrm W}) =  0.0043$, 
$E_{\mathrm PNC} (\delta D) = 0.0035$, and $E_{\mathrm PNC} (\delta E) = -0.0028$. 
Thus the lowest-order corrections given in Eq.~(\ref{Eqn_corr_HF}) are increased. 

To improve the consistency  of the calculation,  one can combine all-order Coulomb-correlated 
matrix elements and experimental energy denominators tabulated in Ref.~\cite{Blundell_PNC}  with
the present third-order Breit corrections. The results are:
\begin{eqnarray}
E_{\mathrm PNC} (\delta H_{\mathrm W}) &=&  0.0047 \; (0.5\%) \, , \nonumber \\
E_{\mathrm PNC} (\delta D) &=&    0.0037          \;  (0.4\%) \, . \label{Eqn_corr_III}
\end{eqnarray}
The Breit correction in energy-denominators  $E_{\mathrm PNC} (\delta E)$ 
was set to zero  because the experimental energies were extensively used in 
Ref.~\cite{Dzuba_PL89,Blundell_PNC}. For example, the experimental energies
were employed in eight out of ten  test cases in the scatter analysis of 
Ref.~\cite{Blundell_PNC} based on Eq.~(\ref{Eqn_E_PNC}).  
The total 0.9\% Breit correction, $\delta E_{\mathrm PNC} =0.0084$, is two times larger
than the corresponding lowest-order modification, which is rather common 
in conventional Coulomb-correlated calculations.
An even larger 2\% Breit correction was found in related  
calculations of the electric-dipole-moment enhancement factor in thallium~\cite{Lindroth_Breit}.
  
{\em Discussion ---}
Combining the calculated 0.9\% Breit correction
with the reference Coulomb-correlated value, Eq.(\ref{Eqn_refVal}), one obtains the 
parity-nonconserving amplitude
\[
 E^{C+B}_{\mathrm PNC}(^{133}{\rm Cs} ) = -0.8991(36) \times 10^{-11} i(-Q_{\rm W}/N) \, .
\]
A 0.4\% theoretical uncertainty is assigned to the above result 
following the analysis of Ref.~\cite{BennettWieman_PRL99}. 
Since the Breit  interaction contributes at the  0.9\% level 
to the total PNC amplitude, even a conservative 10\% uncertainty 
in $\delta E_{\mathrm PNC}$ barely affects the accuracy of $E_{\rm PNC}$. 
When  $E^{C+B}_{\mathrm PNC}$ is  combined with  the experimental values of 
the transition 
polarizability $\beta$~\cite{BennettWieman_PRL99} and $E_{\rm PNC}/\beta$~\cite{Wood_PNC97},
one obtains for the weak charge:
\[
Q_{\rm W}(^{133}{\rm Cs} ) = -72.65(28)_{\rm expt}(34)_{\rm theor} \, .
\]
This value differs from the prediction~\cite{QW_StdModel} of the 
Standard Model $Q_{\rm W}^{\rm SM}=-73.20(13)$ by 1.2$\sigma$, 
versus 2.5$\sigma$  of Ref.~\cite{BennettWieman_PRL99}, where $\sigma$ is calculated
by taking experimental and theoretical uncertainties in quadrature. 
This 1.2$\sigma$ deviation is slightly reduced further by
taking into account corrections for  the neutron nuclear distribution in $^{133}{\rm Cs}$, 
estimated but not included in the final $E_{\rm PNC}$ of Ref.~\cite{Blundell_PNC}.
Recently Pollock and Welliver~\cite{Pollock99}
determined the relevant modification to be $\Delta Q_{\rm W}^{\rm SM} = +0.11$, 
which reduces the deviation from the Standard Model to 1.0 $\sigma$.  
 

The present calculation also provides 
a large Breit correction to the $6S_{1/2}-7P_{1/2}$ electric-dipole
matrix element. Using the {\em ab initio} all-order Coulomb-correlated value~\cite{Blundell_matels},  
$\langle 6S_{1/2}|| D || 7P_{1/2} \rangle = 0.279$, and adding the  0.7\% Breit correction of 0.0019, one
finds  $\langle 6S_{1/2}|| D || 7P_{1/2} \rangle = 0.281$ in  much better agreement with the
0.284(2) experimental value~\cite{Expt_old}.
The calculated Breit corrections bring most of the 
{\em ab initio} Coulomb-correlated hyperfine-structure constants 
for $^{133}$Cs~\cite{Blundell_matels} 
into 0.1\% agreement with experimental values.

To summarize, third-order many-body calculations of the contribution of the Breit interaction 
to the $^{133}$Cs parity-nonconserving
amplitude $E_{\mathrm PNC}$ and relevant atomic properties are reported. 
The difference between the present and the earlier calculations~\cite{Blundell_PNC}
is due to additional inclusion of two-body Breit interaction, correlations, and the consistent
use of experimental energies. The present analysis is a higher-order extension of my recent 
calculation~\cite{myEprint}. 
Since the major correlation effects are 
included, the present third-order analysis seems sufficient.
The calculations reveal
a 0.9\% correction to $E_{\mathrm PNC}$ leading to a reduction to 1.2$\sigma$ 
of the recently 
reported 2.5$\sigma$ deviation~\cite{BennettWieman_PRL99} of the $^{133}$Cs weak
charge from the Standard Model value. If corrections for the  neutron distribution 
in $^{133}$Cs nucleus are included,
then the agreement between the atomic PNC in $^{133}$Cs  and the Standard Model stands at 1.0$\sigma$.
Thus the result reported here brings the most accurate atomic PNC measurement to date~\cite{Wood_PNC97}
into substantial agreement with the Standard Model.

This work was supported by the U.S. Department of Energy,
Division of Chemical Sciences, Office of Energy Research. 
Part of the work has been performed at Notre Dame University during a visit 
supported by NSF grant No.\ PHY-99-70666.
Calculations were partially based on codes developed 
by Notre Dame group led by W.R. Johnson.
The author is thankful to W.R. Johnson and M.S. Safronova 
for useful discussions and H.R. Sadeghpour for suggestions on the manuscript.
Help with the manuscript and the stimulating interest of R.L. Walsworth 
is greatly appreciated.

\begin{table}
\caption{Breit corrections to matrix elements and energy denominators
in a.u.; $\delta X, \rm{I} \equiv X_{\rm BDHF} - X_{\rm DHF}$, and 
$\delta X, \rm{I+II+III}$ are the differences in the third order of MBPT. 
\label{Tab_HF}}  
\begin{tabular}{lrrrr}
&
\multicolumn{1}{c}{$6S_{1/2}-6P_{1/2}$ } & 
\multicolumn{1}{c}{$6S_{1/2}-7P_{1/2}$} & 
\multicolumn{1}{c}{$7S_{1/2}-6P_{1/2}$} & 
\multicolumn{1}{c}{$7S_{1/2}-7P_{1/2}$} \\
\hline           
$ H_{\rm W}$, DHF       &   0.03159  &  0.01891 &  0.01656  &  0.009913 \\
$\delta H_{\rm W}$, I    & -0.00010  & -0.00006 & -0.00005  & -0.000031 \\
$\delta H_{\rm W}$, 
I+II+III                &  -0.00028  & -0.00016 & -0.00014  & -0.000084  \\[3pt]
$ D $,      DHF          &   2.1546   &  0.15176 &  1.8017   &  4.4944   \\
$\delta D$, I            &   0.0001   &  0.00073 &  0.0019   & -0.0004   \\
$\delta D$, I+II+III     &  -0.0004   &  0.00077 &  0.0020   & -0.0012   \\[3pt]
$ \Delta E  $, DHF       &  -0.041752 & -0.085347&  0.030429 & -0.013166  \\
$ \delta \Delta E$, I    &  -0.000020 &  0.000003& -0.000030 & -0.000007  \\  
$ \delta \Delta E$, I+II &  -0.000045 & -0.000023& -0.000034 & -0.000012  \\
\end{tabular}
\end{table}

\begin{references}
\bibitem{stretch_red} $D$ are ``stretched'' amplitudes related 
to reduced  matrix elements as
\mbox{ $\langle nS_{1/2}| D | n'P_{1/2} \rangle = 
 \langle nS_{1/2}|| D || n'P_{1/2} \rangle/\sqrt{6}$}.
\bibitem{Blundell_comment} The Breit correction in Ref.~\cite{Blundell_PNC} also
contains  small random-phase-approximation diagrams in the Coulomb interaction
for  matrix elements of $H_{\rm W}$. The two-body Breit interaction has
been disregarded in Ref.~\cite{Blundell_PNC}.
\bibitem{MBPTcomment} The  random-phase-approximation sequence is limited to  third order; 
and in addition only Brueckner-orbital corrections  are included in the third order calculations.
\end{references}
\end{document}